\documentstyle[aps,preprint]{revtex}
\begin{document}
\draft

\title{Classical and Quantum Chaotic Behaviors of
Two Colliding Harmonic Oscillators
}
\author{Qing-Rong Zheng$^{1,2}$, Gang Su$^{1,3}$ and De-Hai Zhang$^{1,4}$
}
\address{
$^1$ Physics Department, Graduate School, Chinese Academy of
Sciences, P.O. Box 3908,
Beijing 100039, China \\
$^2$ Department of Physics, The University of Hong Kong, Pokfulam Road, Hong
Kong\\
$^3$ Institut f\"ur Theoretische Physik, Universit\"at zu K\"oln, Zulpicher
Strasse 77, D-50937 K\"oln, Germany\\
$^4$ Center of Theoretical Physics, CCAST (World Laboratory),
P.O. Box 8730, Beijing 100080, China
}
\maketitle
\begin{abstract}
We have systematically studied both classical and quantum
chaotic behaviors of two colliding harmonic oscillators.
The classical case falls in Kolmogorov-Arnold-Moser class. It is shown that
there exists an energy threshold, above which the system becomes
nonintegrable.
For some values of the initial energy near the threshold,
we have found that the ratio
of frequencies of the two oscillators affects the Poincar\'e sections
significantly. The largest Lyapunov character exponent depends
linearly on the ratio of frequencies of the two oscillators away from the
energy threshold in some chaotic regions, which shows that the chaotic
behaviors of the system are mainly determined by the ratio.
In the quantum case, for certain parameters, the distribution of the
energy level spacings also varies with the
ratio of frequencies of the two oscillators. The relation between
the energy spectra and the ratio of frequencies of the two oscillators, the
interaction constant, and the semi-classical quantization constant, is also
investigated respectively.
\end{abstract}

\section{Introduction}

Many simple systems show rather complicated classical and quantum chaotic
behaviors due to collision. A well-known example is the
billiard system which describes the collision of a free mass point with
various shapes of boundaries$[1]$. Another example is the occurrence of the
chaotic scattering due to collisions$[2]$. We know that there are many
processes related to the collision phenomena in the real world, such as
chemical reactions$[3]$, celestial mechanics$[4]$, charged particles in
an electromagnetic field$[5]$, processes in hydrodynamics$[6]$, scattering in
atomic physics$[7]$, and transportation problems in mesoscopic physics$[8]$,
etc. A lot of papers are devoted to investigate collision problems in
literature.
The study of colliding phenomena, however, is far from complete, even for the
elastic collision.

The elastic collision can be assumed to be described by a so-called point
contact
interaction or $\delta $-interaction. In the regime of quantum mechanics,
several one-dimensional systems with $\delta $ -interaction potential are
exactly solved$[9,10,11]$. Especially, the solutions of many body systems
with $\delta$-interaction have deep influences on
physics even on mathematics$[10,11]$. A lot of people have studied the
following Hamiltonian%
$$
H=\sum_i\frac{1}{2m} P_i^2+\sum_{ij}\lambda _{ij}\delta (x_i-x_j),\eqno(1)
$$
with $P_i$ the momentum of the ith particle. A simple generalization of eq.(1)
is that each
particle has its own potential $V(x)$ such that%
$$
H=\sum_i(\frac{1}{2m}P_i^2+V_i(x_i))\;+\;\sum_{ij}\lambda _{ij}\delta
(x_i-x_j).%
\eqno(2)
$$
This Hamiltonian system, in general, can be exactly solved neither
in classical mechanics nor in quantum mechanics. The system of two colliding
harmonic oscillators is just the nontrivial reduction of the Hamiltonian
eq.(2).

In this paper, we will study systematically, both in the sense of classical
mechanics and quantum mechanics, the simplest nontrivial reduction of the
extended Hamiltonian eq.(2): two colliding harmonic oscillators, i.e.,  two
harmonic oscillators with $\delta $-interaction. This system is
conserved, and the classical case falls in Kolmogorov-Arnold-Moser(KAM)
class$[12]$. Besides, there exists an energy threshold, above which the
system becomes nonintegrable. For some values of the initial energy, we have
found that the ratio of frequencies of the two oscillators affects the
Poincar\'e sections significantly. The largest Lyapunov character
exponent(LCE)  depends linearly on the ratio of frequencies of the two
oscillators away from the energy threshold in some chaotic regions.
In the quantum case, we have studied the distribution of the energy level
spacings and the density of probability determined by the wave function,
and the
relation between the energy spectra and the ratio of frequencies of the
two oscillators,  the interaction constant and the semi-classical
quantization
constant, respectively. The paper is organized as follows:
The classical case of the two
colliding oscillators will be discussed in Sec. II, and the quantum case
will be studied in detail in Sec. III. The discussions and concluding
remarks will be given finally.

\vspace{0.3cm}

\section{The Classical Case}
As stated above, we will study the two colliding oscillators,
 the simplest nontrivial reduction of the
Hamiltonian  eq.(2). The Hamiltonian of the system can be written as
$$
H=H_1+H_2+H^{\prime },\eqno(3)
$$
where
$$
H_1=\frac{P_1^2}{2m_1}+\frac 12k_1(x_1+b)^2,\eqno(4)
$$
$$
H_2=\frac{P_2^2}{2m_2}+\frac 12k_2(x_2-b)^2,\eqno(5)
$$
and $H^{\prime }$ is the point contact interaction%
$$
H^{\prime }=V\;\delta (x_1-x_2),\eqno(6)
$$
where $k_1$, $k_2$ $>0$ are stiffness constants, $2 b$ is the distance
between two  oscillators, and $V$ is the interacting constant.

The classical dynamic equations of the system can be in principle written
down, but the
resulting equations contain the first order derivative of $\delta
$-function explicitly.
Since numerical methods can not deal with such an indefinite quantity
directly, an approximate expression of $\delta $-function must be
used in the situation. Whereas in the quantum case it is not  so, i.e.,
 the $\delta $-function can be treated exactly. Based on this consideration
and for the consistency of the two cases, we will use the collision
condition to replace the $\delta $-interaction in classical case.
Such a replacement is equivalent to the case that the $\delta$-function
is directly used.
Therefore, the equations of motion of our system are written down
as follows%
$$
\left\{
\begin{array}{ll}
\dot x_1= & P_1/m_1 \\
\dot P_1= & -k_1(x_1+b) \\
\dot x_2= & P_2/m_2 \\
\dot P_2= & -k_2(x_2-b).
\end{array}
\right. \eqno(7)
$$
for $x_1 \neq x_2$, and for $x_1 = x_2$ with an additional condition: the
exchange
of $P_1$ and $P_2$ at $x_1 = x_2$. In this way, it is easy to understand that
the behaviors of the system are
independent of the nonzero parameter $V$.
For simplicity, we will take the values of parameters as%
$$
b=1,\;\;m_1=m_2=1,\;\;\mbox{and}\;\;k_2=1\eqno(8)
$$
in subsequent discussions.

If the two oscillators, with fixed total energy, are separated larger
enough, apparently, there might be no collision at all. Hence, there should
exist an energy
threshold $E_c$ . For a given energy $E$, the system is integrable
as $E<E_c$,  and the system is in general  nonintegrable as $E>E_c$.
The energy threshold corresponds to the case of $x_1=x_2=x$ and $P_1=P_2=0$,
at which the collision just takes place,  without exchanging the
momentum. We know that the energy in this case can be written as
 a function of colliding position $x$ and the separation parameter $b=1$:%
$$
E(x,b=1)=\frac 12k_1(x+1)^2+\frac 12k_2(x-1)^2.\eqno(9)
$$
For the parameter $b=1$ fixed, $E(x,1)$ has a minimum value with respect to
$x$ = $x_c$. We thus have
$$
k_1(x_c+1)+k_2(x_c-1)=0.\eqno(10)
$$
$x_c$ can be obtained
$$
x_c=\frac{(k_2-k_1)}{k_1+k_2}.\eqno(11)
$$
Therefore, substituting $x_c$ into eq.(9), we have%
$$
E_c\equiv E(x_c,1)=\frac{2k_1}{k_1+1}~~<~~2,\eqno(12)
$$
where the conditions in eq.(8) are used. The system will be nonintegrable if
the initial energy $E \ge 2$ for any $k_1$.  This result is
in agreement with the conclusion given in Ref. $[13]$.
We will only consider the case for $E \ge  2$.

To investigate the classical behaviors of the system, it is needed to obtain
the corresponding Poincar\'e sections. To realize this,  we have to integrate
 out  eqs. (7) numerically. Here it should be noted that the ratio of
frequencies of the two oscillators is related to the ratio  $k_1/k_2$
by
$$
\frac{\omega _1}{\omega _2}=\sqrt{\frac{k_1}{k_2}}.\eqno(13)
$$
One may infer that the ratio $\omega_1/\omega_2$  is rational or irrational,
will give  rise to different dynamic behaviors, or equivalently,
different shapes of the Poincar\'e sections. For this   purpose, let us now
present our numerical results. By integrating out eqs.(7) numerically, we have
 obtained the Poincar\'e sections for $k_1/k_2=3,4,8,9$ with $E=2$ and $5$, and
$k_1/k_2=3,4$ with $E=10$ and $30$, respectively. In the calculation, we have
kept the energy of the system to at least six effective digits. The Poincar\'e
sections   for $k_1/k_2=3,4$ and $E=2$ are shown in Fig. 1(a) and (b). Notice
that in the calculation we have taken 60 different initial values, in the same
region, for both $k_1/k_2=3$ and $4$. It is clear that the former shows some
chaotic behaviors, while the latter shows the quasiperiodic behaviors,
at least for the present initial values. This seems to exhibit that the ratio
of $\omega_1/\omega_2$ is integer or irrational, indeed results in different
dynamic behaviors. However, further
investigation shows that when the energy becomes larger, the differences
between the integer ratios and the irrational ratios are vague. To show this,
 the Poincar\'e sections for $k_1/k_2=3,4$ and $E=5$ are given in Fig.1(c) and
(d) respectively.
Unlike the cases of $E=2$, the differences between the Poincar\'e
sections are ambiguous for $E=5$, i.e., the ratio of $\omega_1/\omega_2$ is
integer or irrational, may not affect significantly the dynamic behaviors of
the two colliding oscillators as the energy becomes larger.
To verify above
statement, we have still calculated the Poincar\'e sections for the other
values of $k_1/k_2$  and $E$.
For instance, the Poincar\'e sections   for
$k_1/k_2=8,9$ and $E=2,5$ are obtained, as shown in Fig.2. One may observe
that it shows the similar behaviors as $k_1/k_2=3,4$.
The Poincar\'e sections for $k_1/k_2=3,4$ and $E=10,30$ are also calculated,
but they show nothing
new except similar behaviors as $E=5$. All above analyses    mean that the
ratio of frequencies of the two oscillators is integer or irrational, will
influence  the dynamic behaviors of the system drastically only when the
energy is near the threshold $E_c$ (notice that $E_c=1.6$ for $k_1=4$ and
$1.8$ for $k_1=9$), i.e., the integer ratio exhibits some regular orbits,
and the irrational  ratio presents some chaotic orbits, as the energy is
near $E_c$.
Whereas the integer or irrational ratio  $\omega_1/\omega_2$
does   not play essential role in the classical dynamic behaviors of the
system, as the energy is far away from the threshold $E_c$, i.e., the
system can exhibit some chaotic behaviors either for the irrational ratio
or for the integer ratio in the case, as far as we have studied.
It should be pointed out  that in above discussion we
usually start from the same initial conditions for given energy $E$.

In addition to, to study the chaotic behaviors of the system in detail,
we have also calculated the largest
Lyapunov character exponent(LCE)$[14]$ for different $k_1/k_2$ and different
energies away from the threshold $E_c$ $(E>2)$ in some chaotic regions.
The result, as shown in Fig.3,
shows that the largest LCE depends on $k_1/k_2$ linearly
$$
\lambda _{\max }=\beta \frac{k_1}{k_2},\eqno(14)
$$
with $\beta $ a constant, and is independent of $E$. This suggests that the
classical chaotic behaviors of the system are mainly controlled by the
ratio of frequencies of the two oscillators, as the initial energy
is away from the threshold.

\section{The Quantum Case}
Now let us study the quantum mechanical behaviors of the two colliding
harmonic oscillators.
The Hamiltonian of the system is
$$
H=H_0+H^{\prime },\eqno(15)
$$
where $H_0=H_1+H_2$, $x_i$, $P_i$ ($i=1,2$) satisfy the canonical commuting
relations, and eq.(7) should be replaced by its corresponding quantum partner.
Evidently, $H_0$ can be exactly solved:
$$
H_0|nm>\;=\;E_{nm}\;|nm>,\eqno(16)
$$
with the eigenfunction
$$
|nm>\;\equiv \;\varphi _n(x_1+1) \varphi _m(x_2-1),\eqno(17)
$$
where
$$
\varphi _l(x_j)=N_l\exp (-\frac{\alpha_j}{2}x_j^2)H_l(\alpha_j x_j),
$$
$$
N_l=\sqrt{\frac{\alpha_j} {\sqrt{\pi
}2^ll!}},\;\;\;\;\;\;\;\;\;\;\;\;\;\alpha_j =
\sqrt{\frac{\sqrt{k_j}}{\hbar} },~~(j=1,2),
$$
and $H_l(x)$ is the Hermite polynomials. $\{|nm>\}$ thus constitutes a
complete orthonormal set. The eigenvalue $E_{nm}$ is
$$
E_{nm}=\hbar \omega _1(n+\frac 12)+\hbar \omega _2(m+\frac
12),~~~n,m=integer\eqno(18)
$$
where $\omega _1=\sqrt{k_1}$ and $\omega _2=\sqrt{k_2}$. However, we must
solve, at present, the
following Schr\"odinger's equation%
$$
H\psi (x_1,x_2)\;=\;E\psi (x_1,x_2),\eqno(19)
$$
with $E$ the eigen energy of the system. To achieve this, we can write
$\psi (x_1,x_2)$ as the linear combination of $|nm>$ :
$$
\psi (x_1,x_2)\;=\sum_{n,m}^Na_{nm}|nm>.\eqno(20)
$$
Because of the orthogonal property of $|nm>$, we have%
$$
\sum_{n,m}^N(H_{nm,lk}-E\delta _{nl}\delta _{mk})a_{nm}=0,\eqno(21)
$$
where
$$
H_{nm,lk}=<nm|H^{\prime }|lk>+E_{nm}\delta _{nl}\delta _{mk},\eqno(22)
$$
and
$$
\delta _{nm}=\left\{
\begin{array}{cc}
0, & \mbox{if }n\neq m \\ 1, & \mbox{if }n=m
\end{array}
\right. .
$$

Since it is difficult to obtain the analytic solution of eq.(21), we
solve it numerically.
The integration in eq.(22) can be carried out by
means of the properties of $\delta $-function and the
Gauss-Hermite integration methods. Note that the matrix which consists
of $H_{nm,lk}$
is real symmetric. To find the energy spectra, the eigen equation must be
 solved.
The maximum number of energy levels we have solved is 961. We find that the
statistical properties of 961 levels are qualitatively same as those of 200
levels. In the following, we will give the numerical results in detail,
in order to show our consequences clearly.

\subsection{$P(S)$ versus $S$ }
The distribution $P(S)$ of the energy level spacings is very interesting. We
take $\hbar ^2=0.2$.
For $k_1/k_2=3$ (The ratio of frequencies is thus an
irrational number), the distribution of the energy level spacings obeys
the Wigner
distribution for different interaction parameter $V$, and exhibits the
energy
repulsion$[15]$. However, for $k_1/k_2=4$ (The ratio of frequencies is an
integer), the distribution of the energy level spacings shows the transition
from Poisson distribution to Wigner distribution if $V$ varies from 0.0001
to 0.1. Fig.4 gives a comparison for the above two cases.
Notice that all figures in Fig.4 are
given for 200 levels.  For $k_1/k_2=8,9$, the similar phenomena occur.
These results can be compared with the case of two
dimensional harmonic oscillator where the $P(S)$ does not exist if the
frequencies are commensurable, and is peaked about a nonzero value of $S$
if the frequencies are incommensurable$[16]$.

\subsection{$\left|\psi(x_1,x_2)\right|^2$ }
The density of probability of the system at $(x_1,x_2)$ plane is given
by $\left| \psi (x_1,x_2)\right| ^2$. Here $\hbar ^2$ is fixed to be $0.2$. The
surface plots
and the related contour plots are given in Fig.5.
The system shows some localization.
It seems that the density of probability is more
local for the case of $k_1/k_2=4$ than  for the case of $k_1/k_2=3$.
The scars in Fig. 5 almost lie in the region which is very near the
accessible region for the classical orbits, but there is no clear
relation between the density of the probability and the classical orbits,
like the case of billiard systems$[17]$. For different $V$ and $k_1/k_2$,
both the eigenvalues and the eigenvectors are so different that it is
difficulty to compare the density of the probabilities for different
parameters. The density of the probabilities has no explicit transition
when the energy (i.e., the eigenvalues) crosses the classical chaotic
energy threshold.

\subsection{$E(V)$ versus $V$}
The relation between the eigenvalues and the interaction constant $V$ is also
studied. $\hbar ^2$ is still fixed to be $0.2$. We find that the eigenvalues
 are only sensitive to
very small values of $V$. When $V$ is larger than $1$, the eigenvalues do not
change with respect to the interaction parameter $V$. When $V$ approaches
zero, there is no degenerate energy level for $k_1/k_2=3$, while there exist
clearly the degenerate energy levels for $k_1/k_2=4$, as shown in Fig.6.
In Fig. 6, except $V=0$, there is no energy level crossing, even
though several lines are very close.  For $k_1/k_2=8,9$, the similar
phenomena occur. We have also calculated the corresponding energy curvature
$[16,17]$
$\langle -\frac{d^2E}{dV^2}\rangle $ numerically where the average is for 50
different levels, as shown in Fig.6. The curvature seems to be independent
of $k_1/k_2$. This result can be fitted by $\frac{a_1}{(1+v^2)^2}$
for small $V$, as given in $[18]$.

\subsection{$E(k_1/k_2)$ versus $k_1/k_2$}
Now let us investigate the relation between the eigenvalues and the ratio
$k_1/k_2$. $\hbar ^2$ is yet fixed to be $0.2$.  $V$ is fixed to be $1$.
The energy repulsion is very clear, as shown in Fig.7, where  $k_1$ takes the
 values from $1$ to $25$.
In Fig.7, there is no energy crossing in the given region  as well.
We have calculated the curvature $\langle -\frac{d^2E}{dk_1^2}\rangle $
($k_2=1$)
numerically where the average is taken for 50 different levels. The result
can also be fitted by $\frac{a_2}{(1+k_1^2)^2}$.

\subsection{$E(\hbar)$ versus $\hbar$ }
The relation between the eigenvalues and the quantization constant $\hbar $ is
studied also. In this case, the parameter $V$ takes two values: $0.001$ and $%
1.0$. The parameter $k_1/k_2$ also takes two values: $3$ and $4$. For $%
k_1/k_2=3$, the energy as the function of $\hbar $ is almost independent of
the values of the parameter $V$. However, for $k_1/k_2=4$, the energy as the
function of $\hbar $ is quite different for different values of the
parameter $V$, as shown in Fig.8.
For $k_1/k_2=4$, when $V$ is small, the degenerate energy level
occurs as $\hbar$ is large, and  when $V$ is large, the degenerate
energy level occurs as $\hbar$ is small. We still can not fully
understand this surprising phenomena for the moment. As expected, the
semiclassical  quantization
constant $\hbar$ plays a crucial role in discussions of the transition from
quantum systems to classical systems.

\section{Discussion and Conclusion}

We have systematically studied both the classical and quantum
chaotic behaviors of two colliding harmonic oscillators. In the classical
case, there exists an energy threshold above which the system can be chaotic.
The system is conserved, and the classical behaviors of the system
belongs to the  KAM class. Moreover, the ratio of the frequencies of two
harmonic oscillators affects the Poincar\'e sections significantly,
as the initial energy is near the threshold. The largest LCE depends
linearly on the ratio $k_1/k_2$ of the two oscillators, and is independent
of the initial energy in some chaotic regions, as the energy is away from
the threshold. This implies that the chaotic behaviors of the system are
mainly controlled by the ratio $k_1/k_2$, not by $E$ in the situation.
In the quantum
case, the ratio of frequencies of the two harmonic oscillators is also
important.
Like in the two dimensional harmonic oscillators, the incommensurable
frequencies play a special role. When we deal with the distribution of the
energy  level spacings, the integer ratio of frequencies shows the transition
from  the Poisson distribution to the Wigner distribution with increasing $V$.
If the
interaction parameter $V$ is not very small, the properties of the system are
independent of $V$.
The relation between the energy levels and the ratio of frequencies of the
two harmonic
oscillators, and the semi-classical quantization constant $\hbar$ is
investigated, respectively. The curvature of the energy levels with
respect to $V$ or
$k_1/k_2$ shows universal properties.
As one notes that, the parameter $V$ plays no role in the classical case,
because the potential barrier at $x_1=x_2$ is infinite, in which makes $V$
playing no role. While in quantum case, there are possibilities for a particle
transmision over an $\delta$-function-type potential barrier, and $V$ should
survive. Our
present results just confirm this fact.
Because of the numerical difficulty, the
total behaviors of the system in the processes of $\hbar$ approaching zero
have not been obtained. Nevertheless,
the partial results have already shown some very interesting
phenomena. One may observe that such an apparently simple system indeed
exhibits rather complicated dynamical behaviors, and a few phenomena can
not be sensibly explained for the moment. This system may yet contain rich
exotic behaviors, and deserves to further study.
How to understand the transition from a microscopic system,
through a mesoscopic system, to a macroscopic system, is still a fascinating
topic.

In summary, for
some values of the initial energy near $E_c$, we have found that the ratio
of frequencies of the two oscillators affects the Poincar\'e sections
significantly. The largest Lyapunov character exponent depends
linearly on the ratio of frequencies of the two oscillators away from the
energy threshold in some chaotic regions, which shows that the chaotic
behaviors of the system are mainly determined by the ratio.
In the quantum case, for certain parameters, the distribution of the
energy level spacings also varies with the
ratio of frequencies of the two oscillators. The relation between
the energy spectra and the ratio of frequencies of the two oscillators, the
interaction constant, and the semi-classical quantization constant, is also
investigated respectively, and some interesting phenomena are presented.

\acknowledgments
 One of authors (GS) would like to thank Prof.
C. N. Yang for drawing his attention to this field and for kind advice.
This work is partially supported by the NSF of China, and by the Alexander von
Humboldt Stiftung.

\vspace{0.5cm}

\noindent{\large References}
\begin{description}

\item[[1]] S. W. McDonald and A. N. Kaufman, Phys. Rev. Lett.
{\bf 42}, 1189(1979); L. A. Bunimovich, Funct. Anal. Appli. {\bf 8},
2254(1974).

\item[[2]] E. Ott and T. T\'el, Chaos, {\bf 3}, 417(1993).

\item[[3]] C. C. Rankin and W. H. Miller, J. Chem. Phys. {\bf 55}, 3150(1971);
B.B. Grayce, R. T. Skodje and J. M. Hutson, J. Chem. Phys. {\bf 98},
3929(1993).

\item[[4]] M. H\'enon, Physica {\bf D33}, 132(1988).

\item[[5]] A. A. Chernikov and G. Schmidt, Chaos  {\bf 3}, 525(1993), and
references therein.

\item[[6]] B. Eckhardt and H. Aref, Philos. Trans. R. Soc. London Ser.
{\bf A 326}, 655(1988).

\item[[7]] J. M. Yuan and Y. Gu, Chaos {\bf 3}, 569(1993).

\item[[8]] H. A. Weidenm\"uller, in {\it Chaos and Quantum Chaos}, ed. by W. D.
Heiss,(Springer-Verlag, Berlin, 1992), p121.

\item[[9]] S. Albeverio, F. Gesztesy, R. Hoegh-Krohn and H. Holden, {\it
Solvable
Models in Quantum Mechanics}, (Springer-Verlag, New York,1988).

\item[[10]] C. N. Yang, Phys. Rev. Lett. {\bf 19}, 1312(1967); C. N. Yang,
Phys. Rev. {\bf 168 }, 1920(1968).

\item[[11]] E. H. Lieb and W. Liniger, Phys. Rev. {\bf 130}, 1605(1963).

\item[[12]] M. C. Gutzwiller, {\it Chaos in Classical and Quantum Mechanics},
(Springer-Verlag, New York,1990), and references therein.

\item[[13]] A.A. Chernikov and G. Schmidt, Phys. Lett. {\bf A 184}, 328 (1994).

\item[[14]] A. Wolf, J. B. Swift, H. L. Swinney and J. A. Vastano, Physica
{\bf D 16}, 285(1985).

\item[[15]] F. Haake, {\it Quantum Signatures of Chaos}, (Springer-Verlag
,Berlin, 1991), and references therein.

\item[[16]] M. V. Berry and M. Tabor, Proc. R. Soc. (London) Ser. {\bf A356},
 375(1977); K. Nakamura, {\it Quantum Chaos}, ( Cambridge Univ. Press, 1993),
and
references therein.

\item[[17]] N. Pomphrey, J. Phys. {\bf B 7},  1909(1974); L. E. Reichl,
{\it The Transition to Chaos}, ( Springer-Verlag, New York, 1992), and
references
therein.

\item[[18]] F. von Oppen, Phys. Rev. Lett. {\bf 73}, 798(1994).

\end{description}

\newpage

{\bf Figure Captions}

Fig.1. The Poincar\'e sections for $k_1=3,4$

Fig.2. The Poincar\'e sections for $k_1=8,9$

Fig.3. The largest Lyapunov character exponent versus $k_1/k_2$($k_2=1$)
 for different energies $E>2$

Fig.4. The distribution of level spacings for different $V$ and $k_1$.

Fig.5. $|\psi(x_1,x_2)|^2$ versus $x_1$, $x_2$  (The wave functions
are not normalized.)

Fig.6. The Energy $E$ and the curvature $\langle -\frac{d^2E}{dV^2}\rangle $
versus the interaction constant $V$

Fig.7.  The energy $E$ and the curvature $\langle -\frac{d^2E}{dk_1^2}
\rangle $ versus $k_1$

Fig.8. The energy $E(\hbar)$ versus $\hbar$

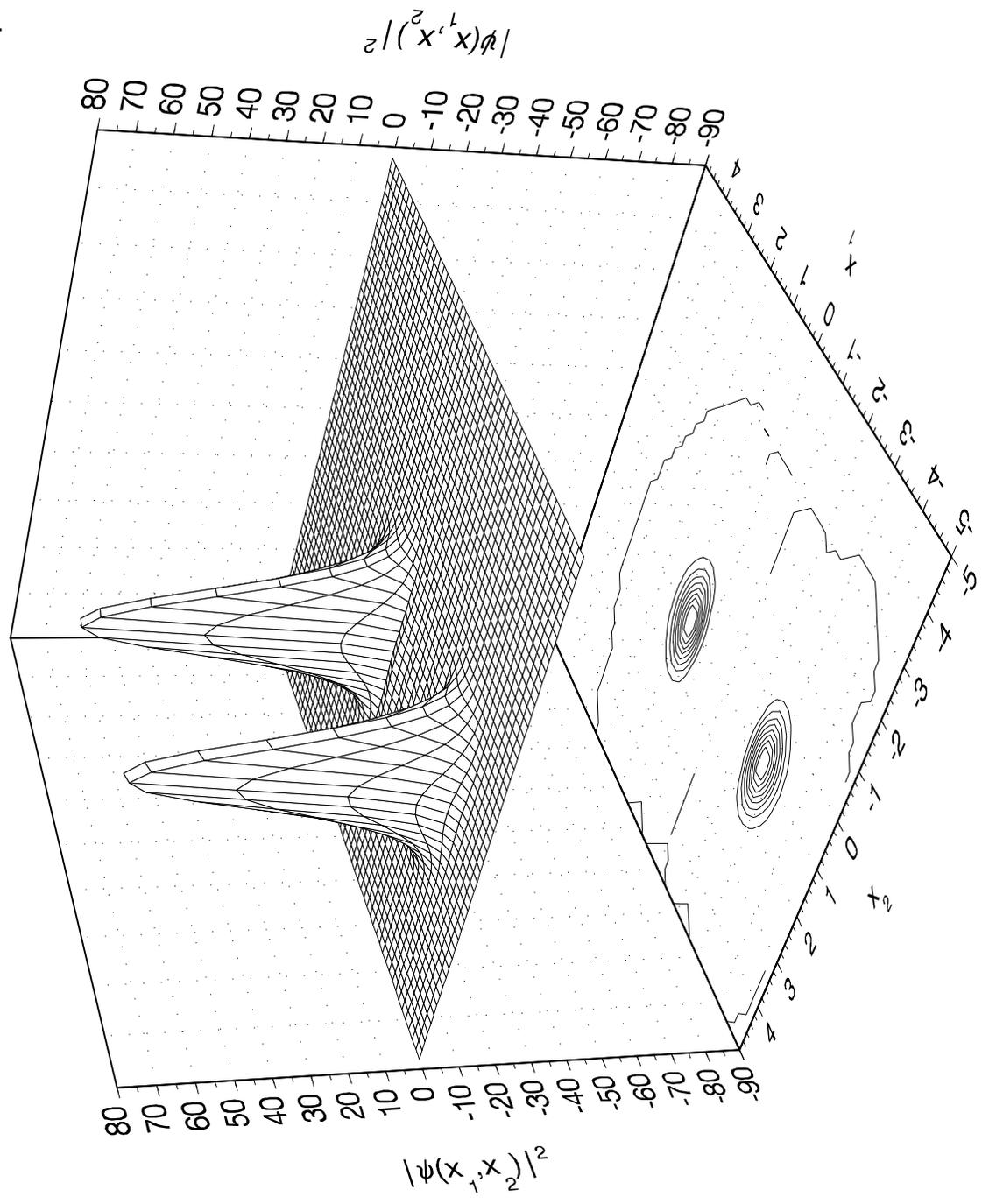
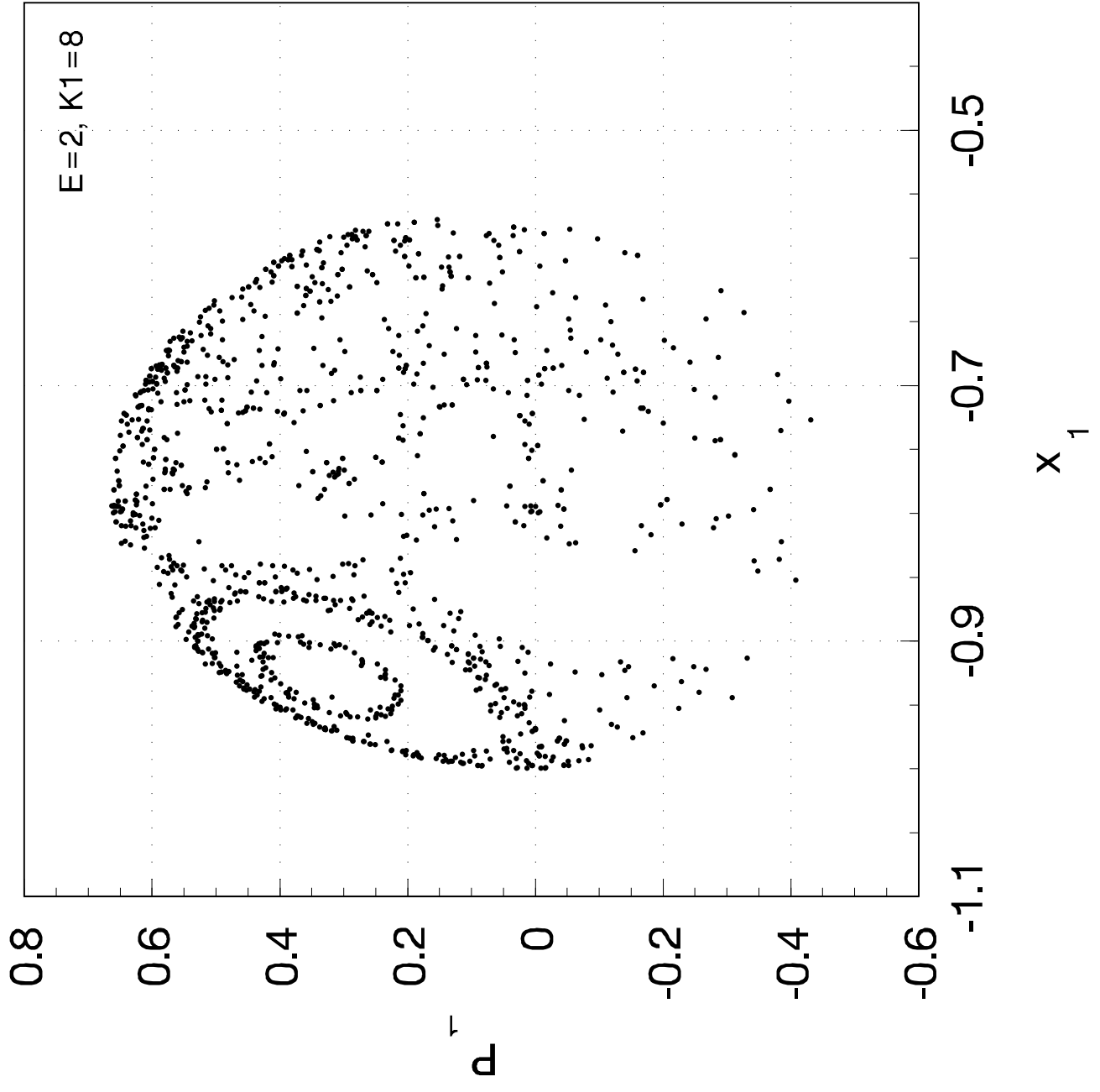
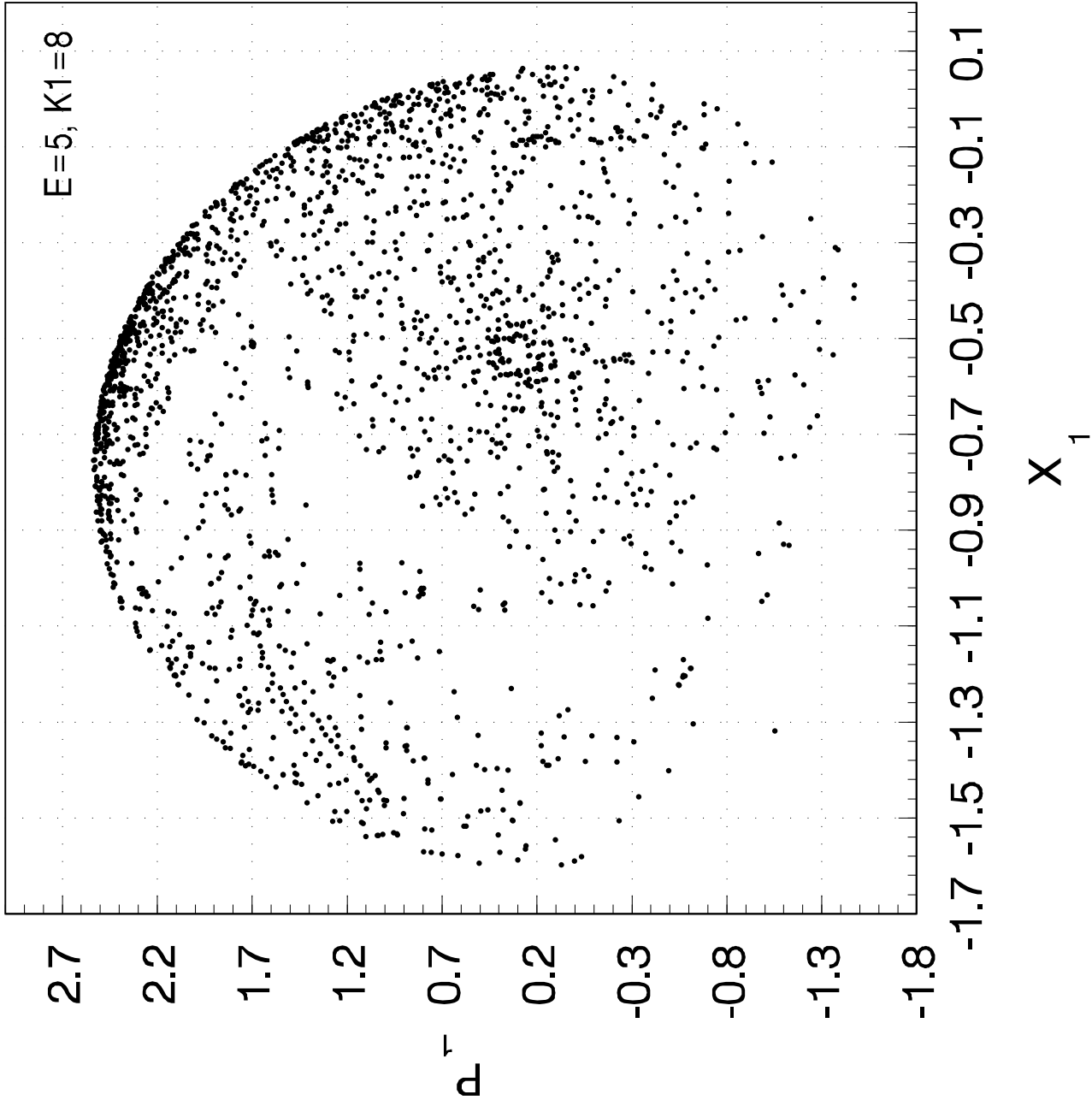
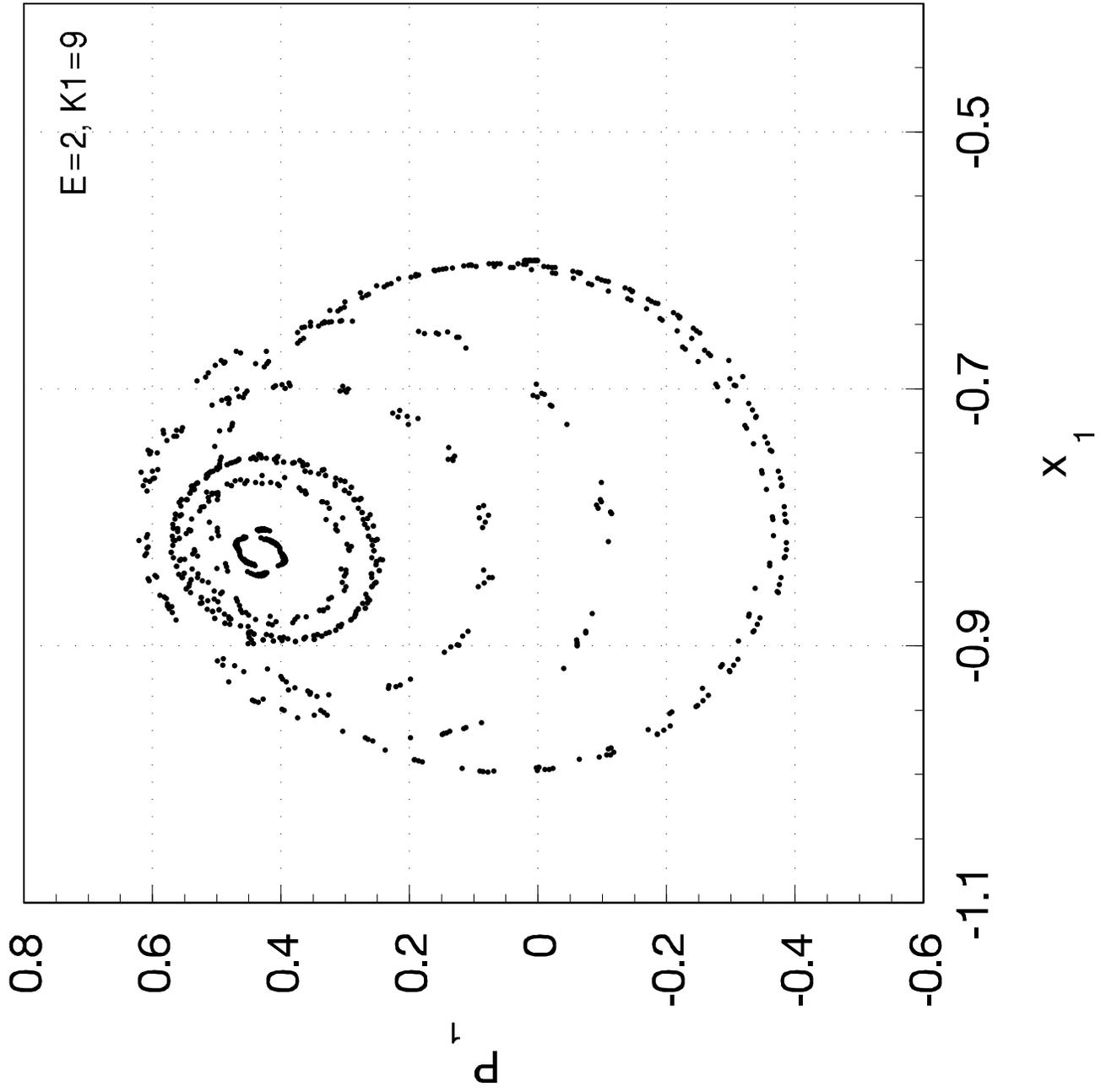
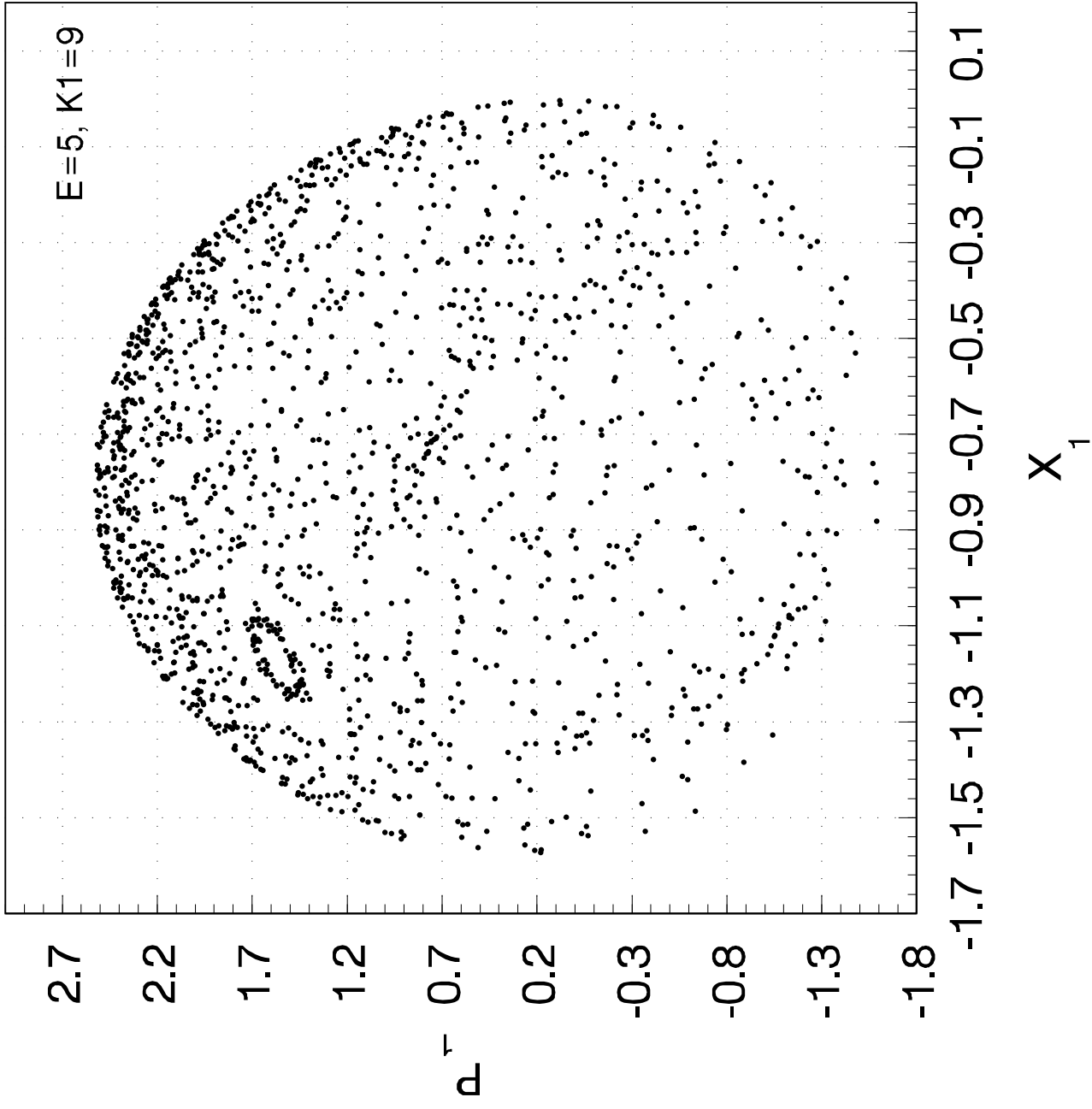
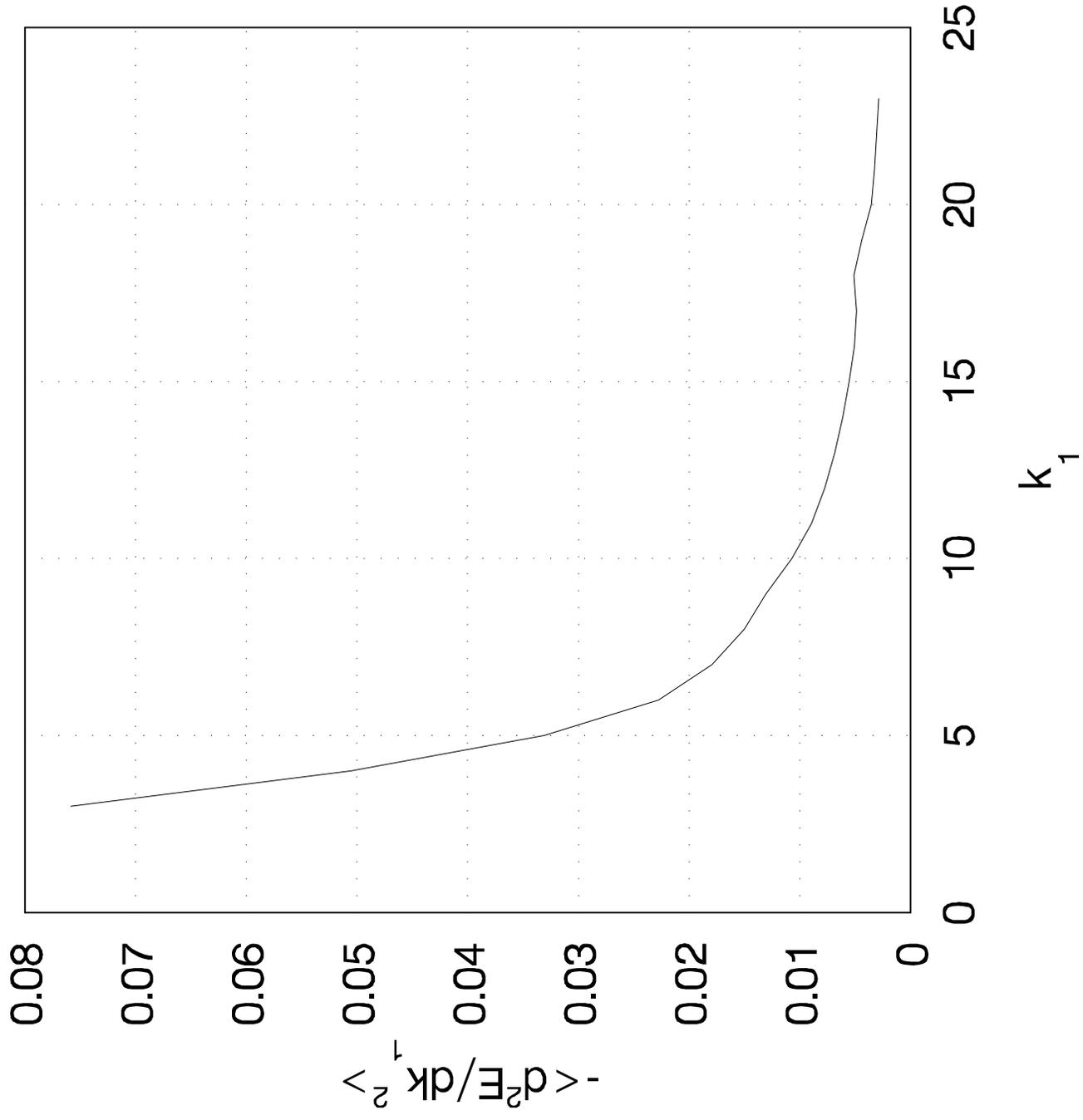
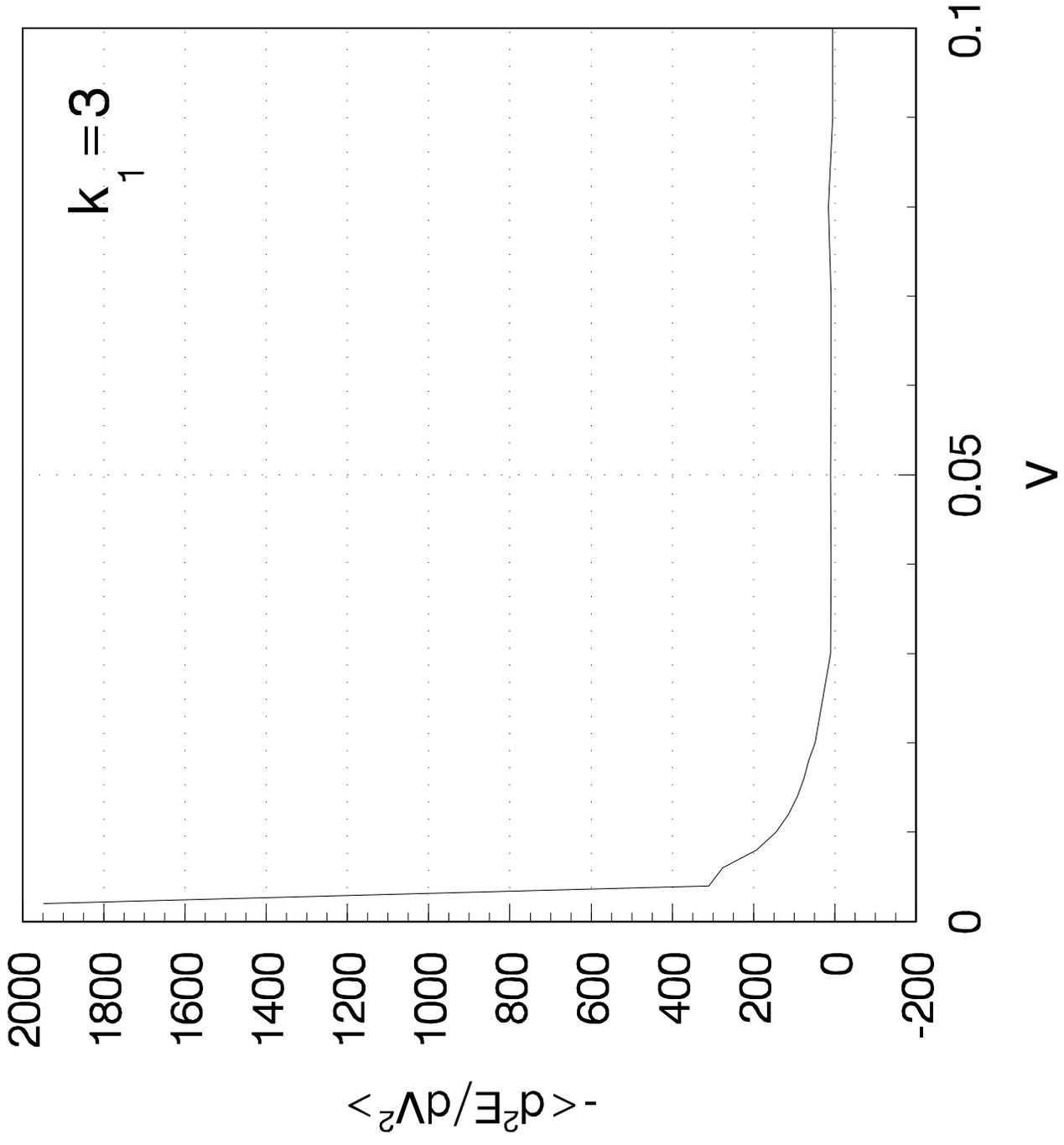
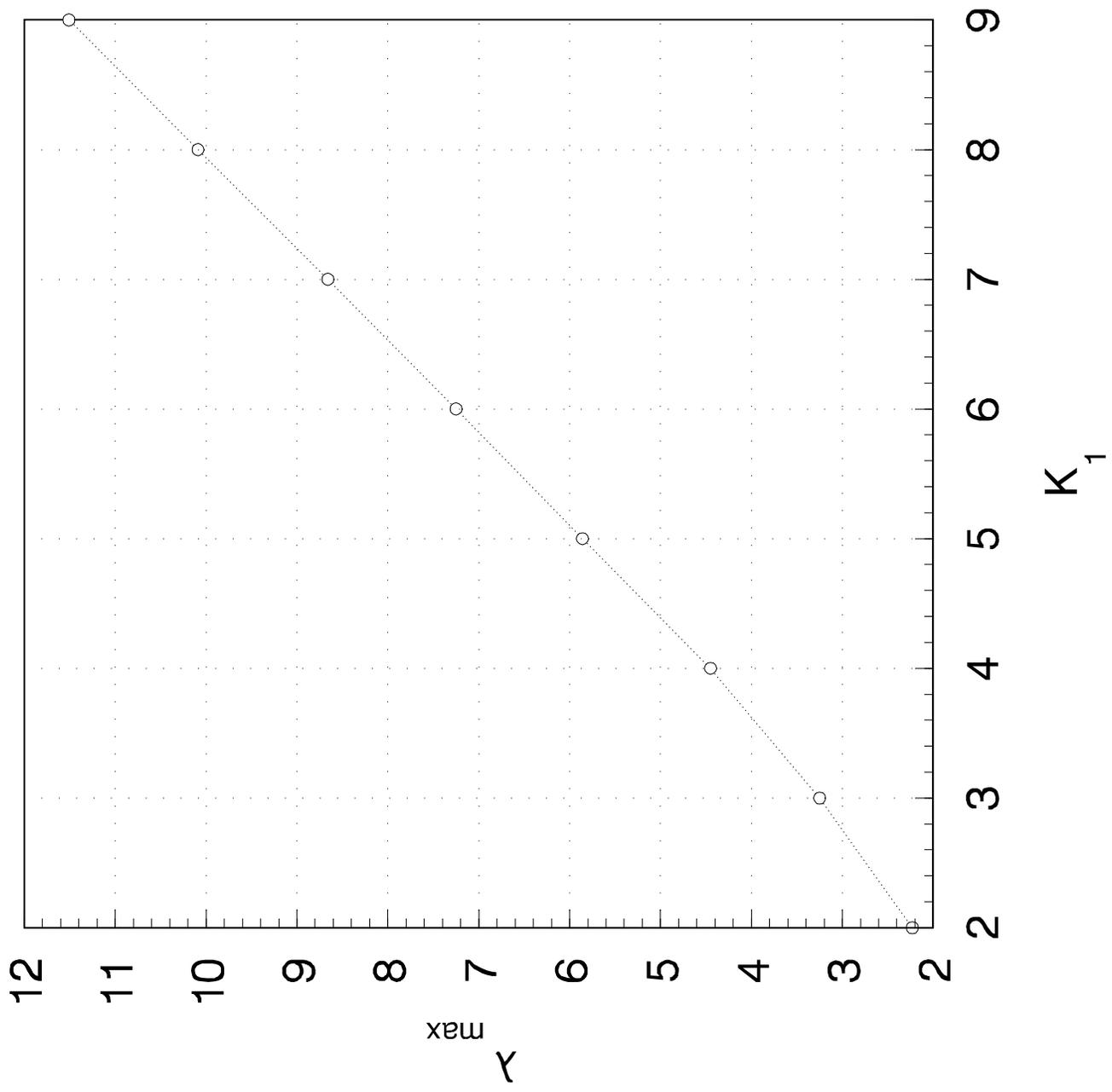
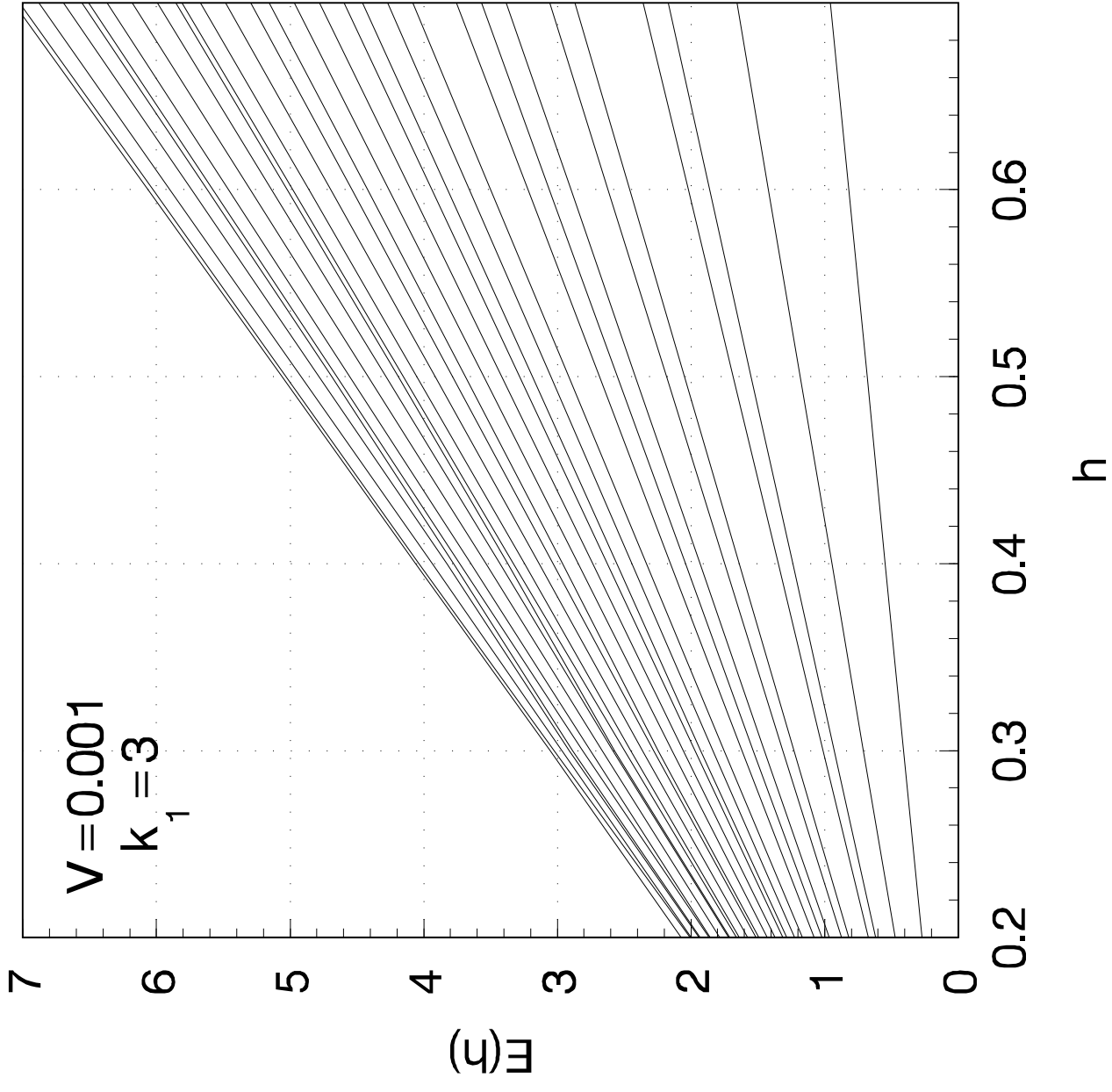
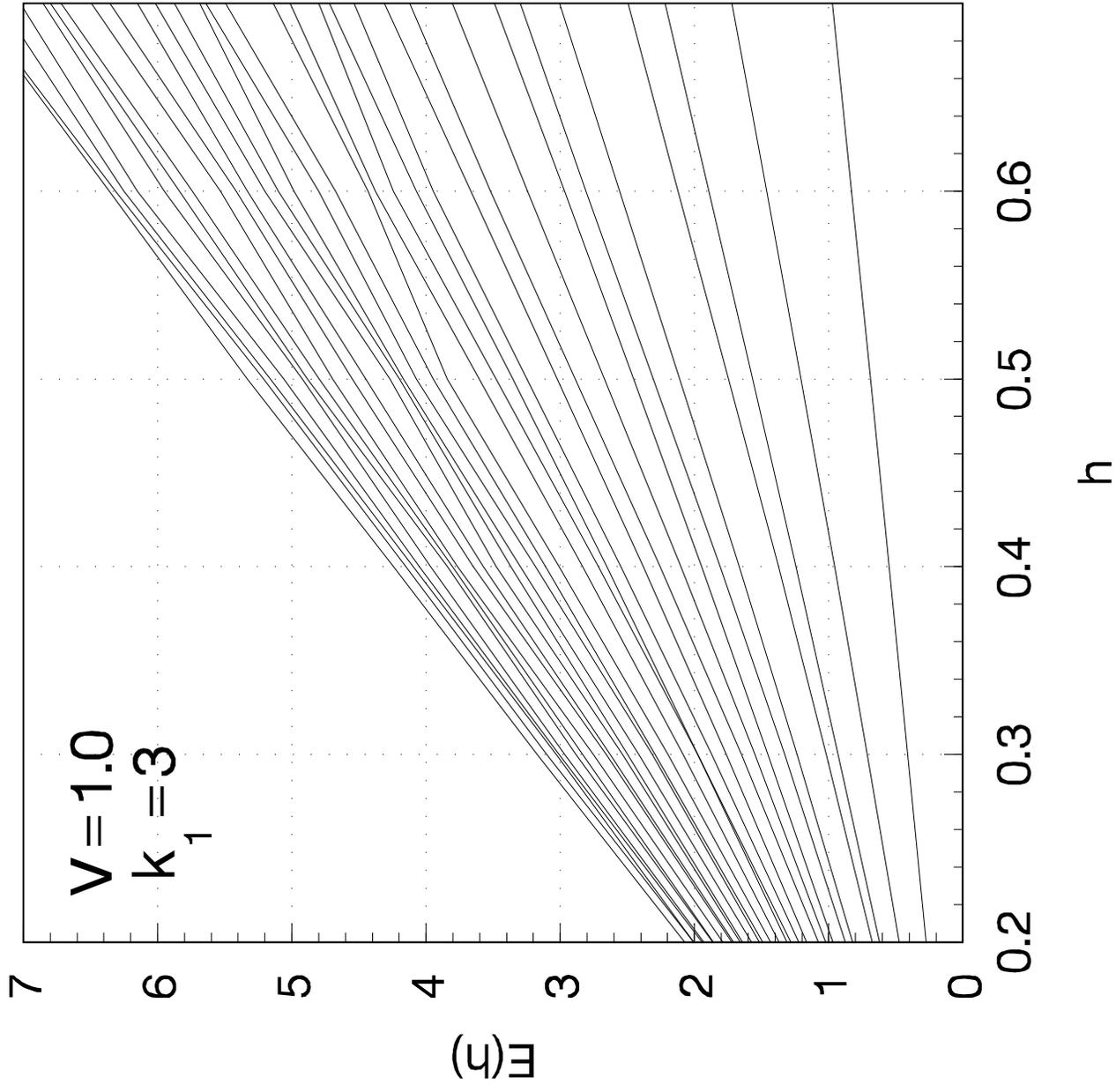
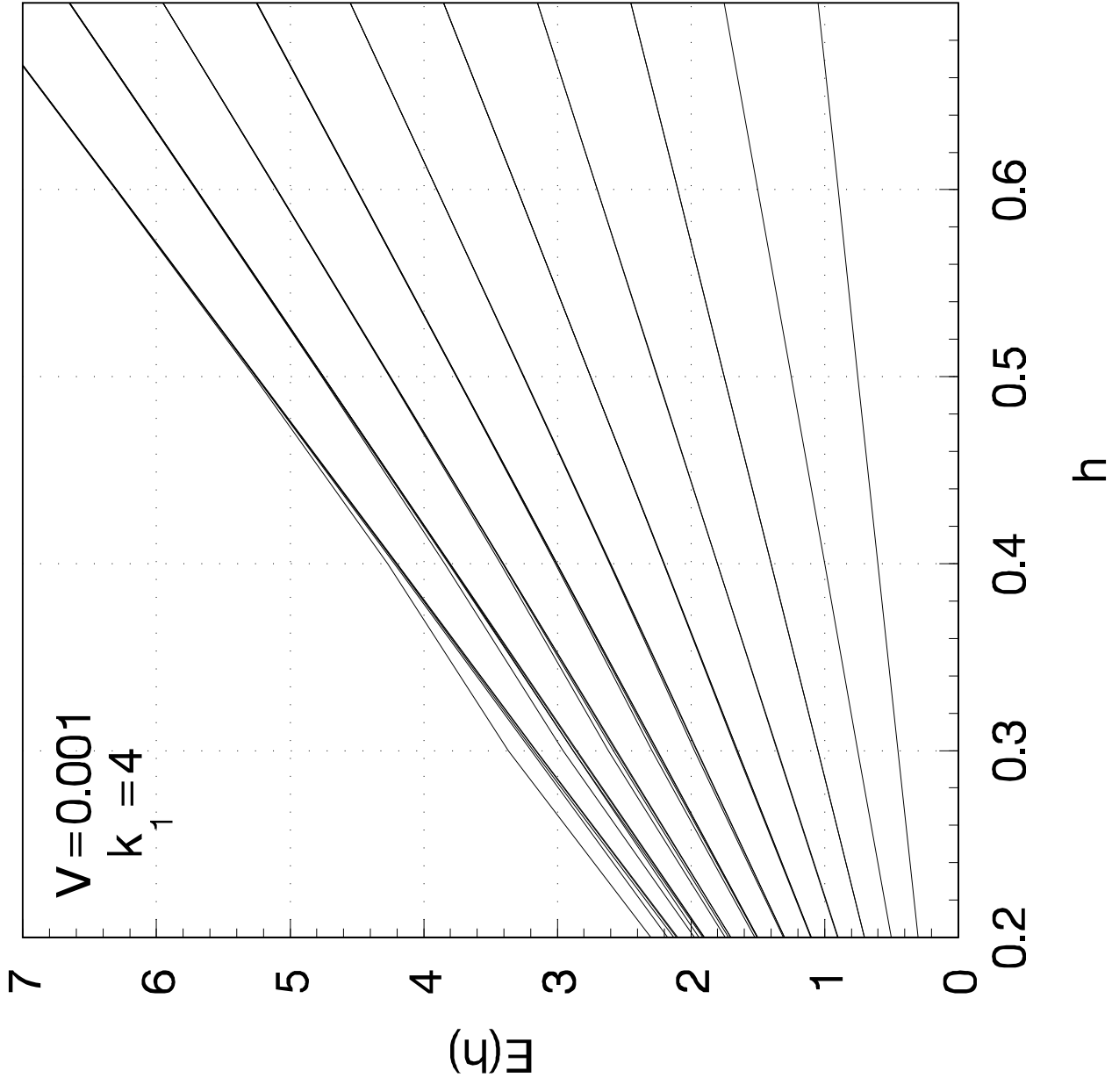
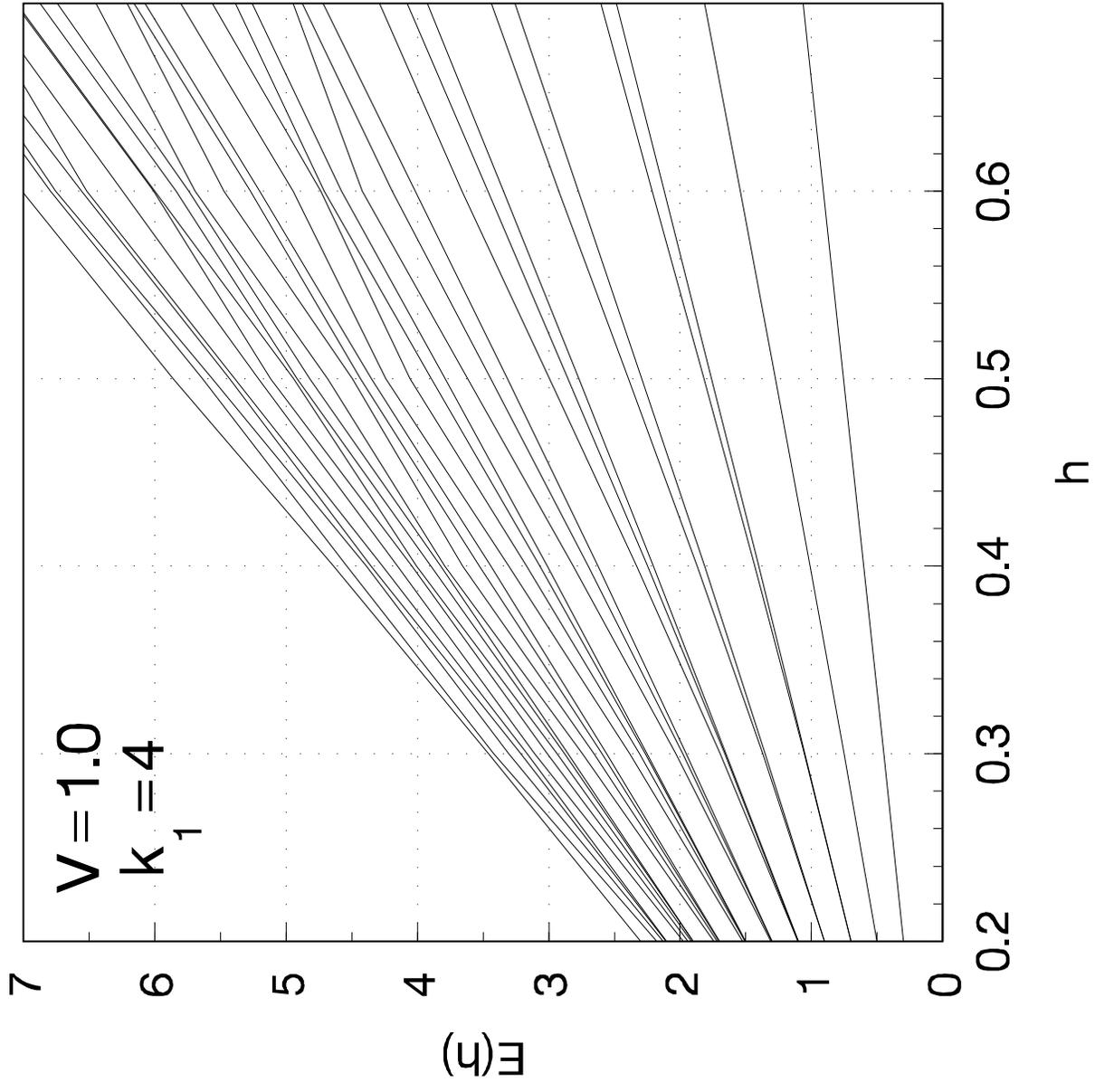
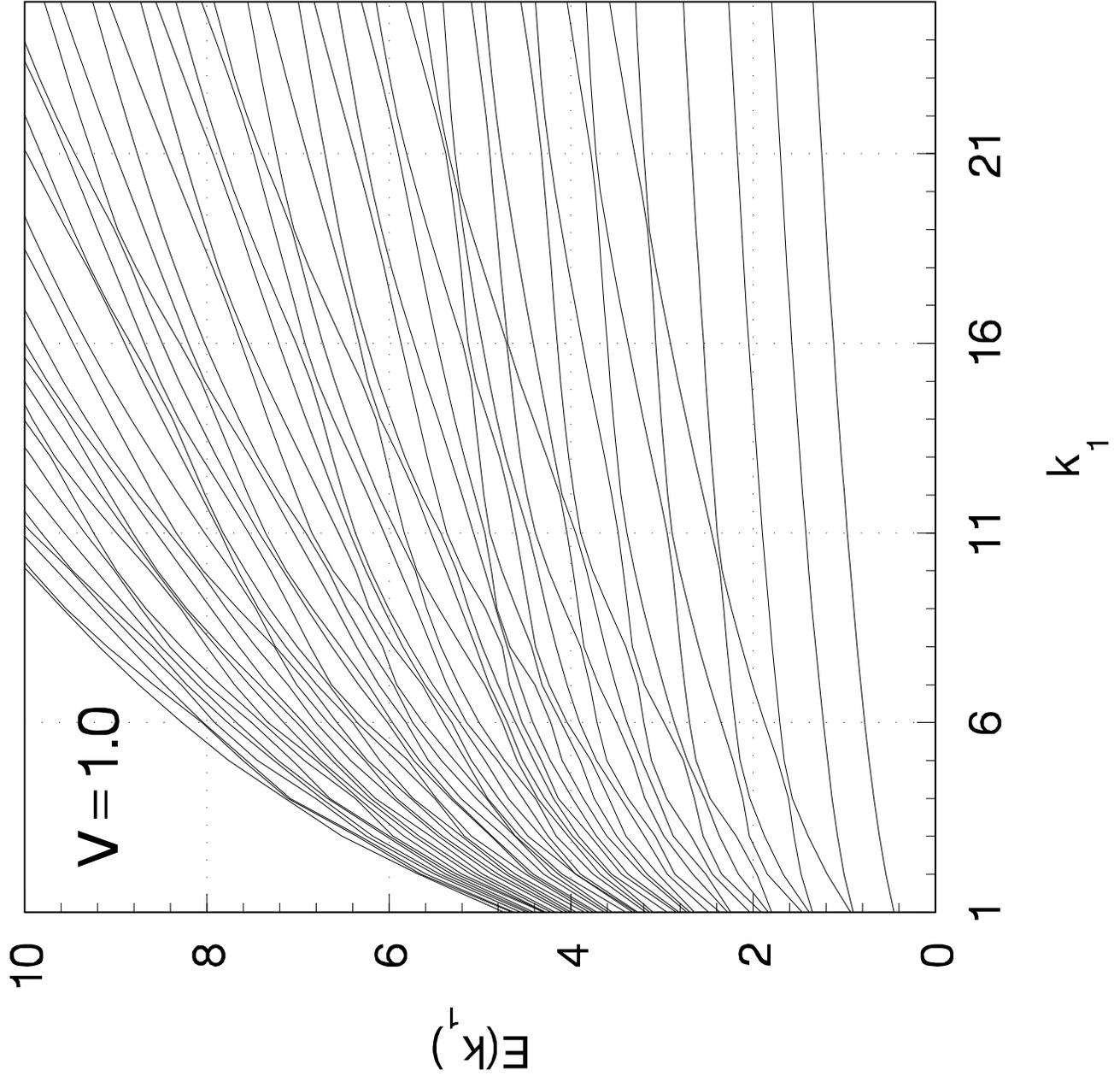
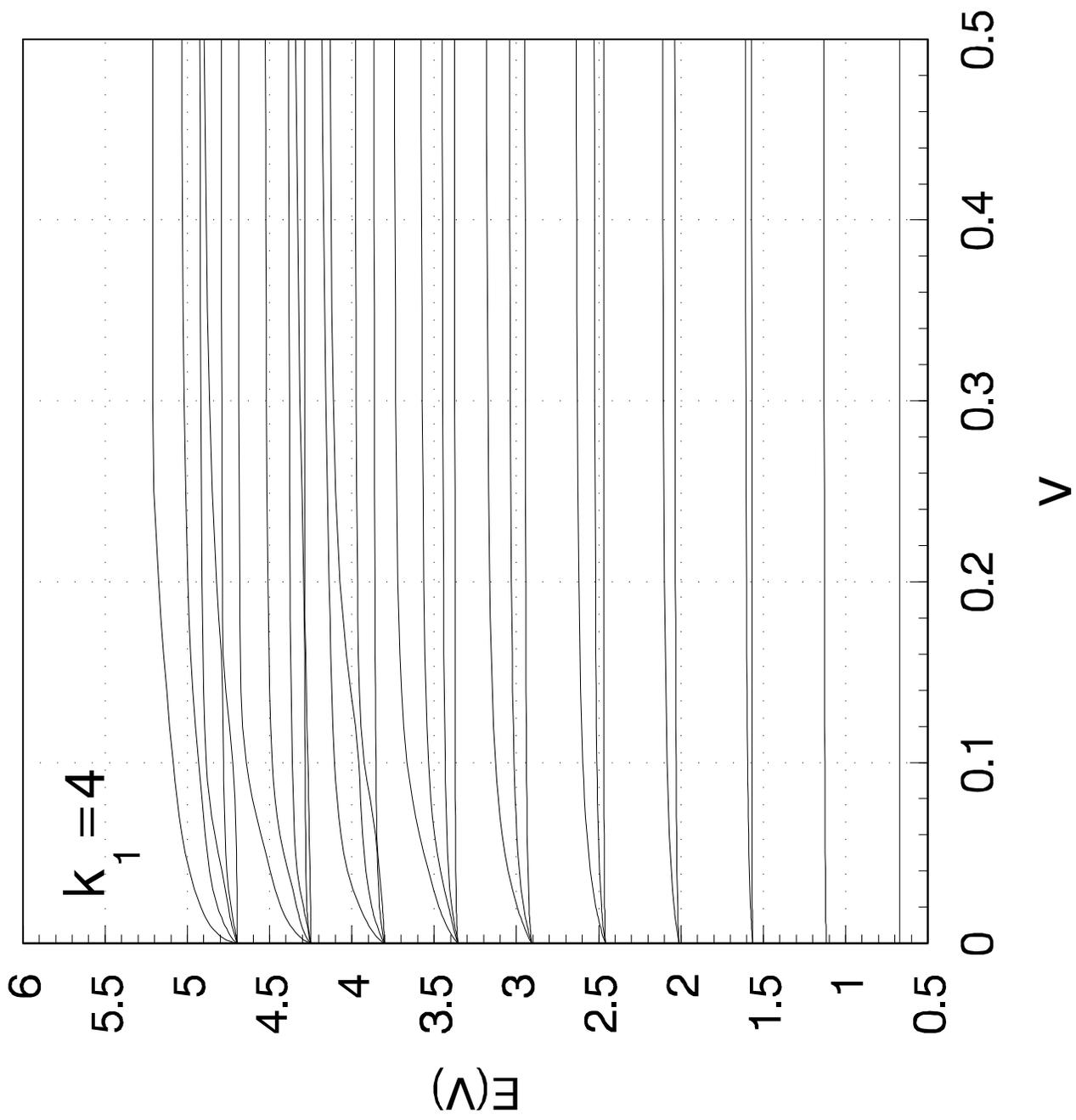
\end{document}